# The modelling of COVID19 pathways sheds light on mechanisms, opportunities and on controversial interpretations of medical treatments. v2

Maria Luisa Chiusano
Department of Agricultural Sciences, University Federico II of Naples
Correspondence to: chiusano@unina.it

**Abstract**

The new coronavirus (2019-nCoV or SARS-CoV2), inducing the current pandemic disease (COVID-19) and causing pneumoniae in humans, is dramatically increasing in epidemic scale since its first appearance in Wuhan, China, in December 2019.

The first infection from epidemic coronaviruses in 2003 fostered the spread of an overwhelming amount of related scientific efforts. The manifold aspects that have been raised, as well as their redundancy offer precious information that has been underexploited and needs to be critically re-evaluated, appropriately used and offered to the whole community, from scientists, to medical doctors, stakeholders and common people. These efforts will favour a holistic view on the comprehension, prevention and development of strategies (pharmacological, clinical etc) as well as common intervention against the new coronavirus spreading.

Here we describe a model that emerged from our analysis that was focused on the Renin Angiotensin System (RAS) and the possible routes linking it to the viral infection. because the infection is mediated by the viral receptor on human cell membranes Angiotensin Converting Enzyme (ACE2), which is a key component in RAS signalling. The model depicts the main pathways determining the disease and the molecular framework for its establishment, and can help to shed light on mechanisms involved in the infection. It promptly gives an answer to some of the controversial, and still open, issues concerning predisposing conditions and medical treatments that protect from or favour the severity of the disease (such as the use of ACE inhibitors or ARBs/sartans), or to the sex related biases in the affected population. The model highlights novel opportunities for further investigations, diagnosis and appropriate intervention to understand and fight COVID19.

**Introduction**

In the past two decades, coronaviruses have caused three epidemic diseases: COVID-19[1], the severe acute respiratory syndrome (SARS)[2], the Middle East respiratory syndrome (MERS)[3].

They are enveloped viruses with a positive sense single-stranded RNA genome (26–32 kb) from sequence analysis revealed a high sequence similarity among the coronaviruses causing the COVID-19, SARS, and MERS[1]. Because of the high similarity with SARS coronavirus, usually termed SARS-CoV, the 2019nCov is also named SARS-CoV-2. The high similarity also favours the understanding of many aspects related of the infection in the human body thanks to knowledge derived from scientific experiments and studies performed on previously human infecting "relatives" of SARS-CoV2[4], which are also causing severe acute respiratory stress.

The genome of SARS-CoV-2 contains at least ten open reading frames (ORFs) useful to viral replication and transcription and coding for structural proteins. Specifically, four main structural proteins have been recognized which are: spike (S), envelope (E), nucleocapsid (N) and membrane (M) proteins. The spike protein plays a critical role to start the viral infection.

**Cell entry of SARS-CoVs**

S proteins of coronaviruses are large membrane linked glycoproteins that are responsible both for binding to receptors on host cells and for membrane fusion[2]. The S proteins are organized by the S1 and S2 domains, which are respectively similar to the S1 and S2 domains in both human immunodeficiency virus (HIV) and influenza proteins. The S1 domain of all characterized coronaviruses, SARS-CoV and SARS-Cov2 included, mediates the initial high-affinity association with their receptors[5]. Experimental analyses proved that Sars-CoV2 does not use known coronavirus receptors, such as aminopeptidase N (APN) and dipeptidyl peptidase 4 (DPP4). SARS-CoV-2, instead, it requires the angiotensin-converting enzyme 2 (ACE2) to enter cells[1]. This was clearly shown for SARS-CoV[5]. Studies on the African green monkey kidney cell line Vero E6, which permits viral replication, highlighted the mechanisms of association of the S1 domain of SARS-CoV with human cell receptors. The authors confirmed that a soluble form of ACE2, but not of the related enzyme termed ACE, blocked association of the S1 domain with Vero E6 cells, and also that anti-ACE2, but not anti-ACE antibodies blocked viral replication in Vero E6 cells.

The role of ACE2 as a functional receptor for SARS-CoV was recently confirmed also for SARS-CoV2[1,4]. It was demonstrated that also SARS-CoV2, as well as SARS-CoV, use endosomal cysteine proteases cathepsin B and L (CatB/L) and the transmembrane serine protease TMPRSS2 for S protein priming in cell lines, a step required for viral entry into the cell.[4] TMPRSS2 had previously been revealed as a host cell factor that is critical for entry in the cells and not for viral replication, affecting the S protein at the cell surface and inducing virus-plasma membrane fusion[6]. Furthermore, lysosome-tropic reagents, like E64D, which inhibits CatB/L, failed to completely inhibit SARS-CoV2, as well as SARS-CoV, entry in the cells. Indeed, the entry full inhibition also needed TMPRSS2 inhibitors (like camostat mesylate)[4]. This concurrent action had been previously shown to be required for robust blockade of SARS-CoV[7].

**Tissue distribution of viral receptors**

Host susceptibility to SARS -CoV2 infection is primarily determined by the affinity between the viral S1 binding site (Receptor Binding Site, RBD) and host ACE2, and this determines the initial viral attachment step[8]. Moreover, since the bind of ACE2 to the S1 domain of S proteins should require the priming of the S protein at the cell surface to induce virus-plasma membrane fusion for viral entry[4], the copresence of both molecule should be investigated to determine the possible occurrence of the infection in the cells.

ACE2 appeared to have a restricted distribution in human tissues, being initially reported to be expressed mainly in the heart, kidney, and testis. [9,10] Interestingly, remarkable findings since 2004[11] revealed the surface expression of ACE2 on lung alveolar epithelial cells and enterocytes of the small intestine, as well as in arterial and venous endothelial cells and arterial smooth muscle cells in all organs studied, explaining the relationship with SARS coronavirus infection. Together with the presence of ACE2 in vascular endothelium, these results highlighted possible routes of entry for the SARS-CoV. Recently, single-cell transcriptomes showed the presence of ACE2 RNA (here indicated as ace2) in type II alveolar cells of lung, esophagus upper and stratified epithelial cells, absorptive enterocytes from ileum and colon, cholangiocytes, myocardial cells, kidney proximal tubule cells, and bladder urothelial cells, suggesting that those organs could be considered as potential high risk for 2019-nCoV infection. In particular, presence of ACE2 was revealed also in the mucosa of the oral cavity, being highly enriched in epithelial cells. Moreover, the research highlighted that among different oral sites, ACE2 expression was higher in tongue than buccal and gingival tissues. Here we also investigated the expression pattern of TMPRSS2 from public biological databases, [https://www.proteinatlas.org/ENSG00000184012-TMPRSS2/tissue]. This analysis revealed that protein and mRNAs are not typically present in lung in physiological condition. Nevertheless, it is interesting to note, that TMPRSS2 expression correlates with SARS-CoV infection in the upper lobe of the lung[6]. The protein is almost present also in paratyroids and salivary glands, while the mRNA expression is detected in lung, salivary gland and in the tongue. These findings further indicate that the mucosa of the oral cavity may be a potentially high risk route for the starting of the infection[12].

It is worthy to mention in this context that chatepsins, that can also contribute to the internalization of the virus[4], are stimulated by cytokines, key proteins in the inflammatory response. Interleukin 6 (IL-6) leads a relevant increase in cathepsin L, whereas TGF-beta1 decreases the amount of cathepsin L, at both mRNA and protein levels, proving that the cytokines IL-6 and TGF-beta 1 modulate cathepsin gene expression.[13] Remarkably, IL6 and not TGF-beta1 appear to by overproduced in COVID-19 patients[14,15].

**Viral infection provokes a cytokine storm in infected patients.**
Public data from documented computed tomography (CT) and radiography imaging from COVID19 infected patients [https://www.sirm.org/category/senza-categoria/covid-19/] highlight that a cytokine storm-syndrome rather than the virus itself could be the cause of severe lung injuries and death.[16]
A cytokine storm is an overproduction of cytokines and an accumulation of immune cells that is often associated with lung inflammation and fluid buildup, and that can lead to respiratory distress and mortality. Moreover, it is widely known (information also available from Wikipedia) that the principal symptoms of a cytokine storm are high fever, fatigue and nausea, which correspond to the principal ones from COVID19 infected patients [https://www.worldometers.info/coronavirus/coronavirus-symptoms/].
Cytokines are proteins produced and secreted by a broad range of cells, including immune cells and endothelial cells, fibroblasts, and various stromal cells in response to varied stimuli. They act as chemical messengers and are released from cells in contact with the stimulus, a pathogen, affecting the actions of other cells by binding to receptors on their surface. Over the past two decades, much has been learned about the diagnosis and treatment of cytokine storm syndromes even for intervention in SARS[17] and MERS[18]. The development of biological therapies for a variety of rheumatic, oncologic, and other conditions, offered novel approaches to treating the immune response, going beyond a not targeted immune suppression by corticosteroids or other relatively non-selective immune suppressants. As an example, recently, a number of specific anti-cytokine approaches have proven effective in treating a variety of cytokine storm syndromes, including those triggered by coronaviruses. Data are becoming available from cytokine content in patients affected by COVID19 as they are from patients involved in previous pandemic infections.[19,20] This reveals the intricate patterning of gene expression associated to the inflammatory status and the possible side effect from untargeted treatments. [17]

**The RAS signalling**
Although its relevance for the SARS-CoVs infection, ACE2[9,10] was first discovered for its role as a transmembrane exopeptidase that catalyses the conversion of angiotensin II (ANG-II) to angiotensin (1-7), a vasodilator[21,22]. Angiotensin II (ANGII) is a peptide hormone that can bind the type 1 angiotensin II receptor (AT1), giving place to a number of effects that result in vasoconstriction and, therefore, in increased blood pressure. ANGII also stimulates the release of aldosterone from the adrenal cortex in the kidney to promote sodium retention. ANGII is produced from the Angiotensin I by the transmembrane protein angiotensin-converting enzyme, or ACE (ACE1), that was discovered in 1956[23]. ACE2 may also convert angiotensin I to angiotensin (1–9)[10], which has a prothrombotic activity[24].
The RAS system acts as a homeostatic regulator of blood pressure, fluid and electrolyte balance, and vascular resistance, as well as of the function of several organs, with a fundamental role in both health and disease. The complexity of the whole system is complicated by its systemic and local tissue-specificity, which includes a wide range of molecules with different biological actions (including the alternative Ang II receptor, the angiotensin type 2 receptor (AT2r), the Mas-related G-protein–coupled receptor, other metabolites, like aldosterone, Ang III, Ang IV, etc.).[25] In a more simplistic way, it is conventional to consider the ACE-ANGII-AT1r and the ACE2-ANG(1-7) pathways as the two main axes representing the Renin Angiotensin System signalling (RAS signalling)[26].

**ACE2 in diseases**
The discovery of an enzyme that cleaves the vasoactive peptide ANGII soon highlighted its possible role in the treatment of hypertension, attracting attention also for pharmacological applications. Several observations and experimental evidence suggested the beneficial role of ACE2 in diseases and cardiovascular disfunctions[27,21]. ACE2 is also essential for the expression of neutral amino acid transporters, influencing the composition of the gut microbiota[28]. ACE2 is shown to increase in a number of pathologies, including myocardial infarction[24], atherosclerosis[29], renal diseases[30,31] and diabetes[32,33], liver cirrhosis[34]. inflammatory lung injuries[35,36] and, in these studies, the general protective role of the enzyme when inhibiting the ANGI-ACE-ANGII axis of RAS is discussed.[37]

The overwhelming scientific production in the field also reports on the spread role of ACE2 increase as anti-inflammatory and anti-oxidant agent, in several tissues, lungs[38], liver[39,40], in neuroprotection[41], and also in atherosclerosis, cerebral ischemia, obesity, chronic kidney disease, and asthma. Indeed, it induces cytokins depletion, antiproliferative effects, anti-fibrosis and anti-hypertrophy effects[42], addressing its key action and its role as target for therapies in inflammatory diseases.[26,43,44,45,46]

ACE2 has been considered as protecting from predisposition to acute respiratory distress syndrome (ARDS)[47], as also demonstrated in SARS-CoV infection, where its decrease, due to the interaction with the virus, favours the syndrome[48]. This indicates that a reduction in pulmonary ACE2 activity also contributes to the pathogenesis of lung inflammation, accompanied by the expression of cytokins that cooperate with the direct effects of viral infection. As an example among the many, the loss of ACE2 function in mouse shows the release of proinflammatory chemokines, such as C-X-C motif chemokine 5 (CXCL5), macrophage inflammatory protein-2 (MIP2), C-X-C motif chemokine 1 (KC), and TNF-α from airway epithelia, increased neutrophil infiltration, and exaggerated lung inflammation and injury[38]. Interesting, smog and nicotins also favour lung inflammation increasing ACE activity and, therefore, affecting ACE2 functionality.[49]

**ACE2 in medical treatments**

The need to reduce the negative effects of the ACE-ANGII-AT1r axis, launched a series of therapeutic approaches aiming to amplify the ACE2-Ang(1–7) axis. ACE inhibitors, AT1R blockers and mineralocorticoid receptor blockers (MRB), pioglitazone and ibuprofen[50] included, are reported to have direct effects on suppression of the ACE-Ang II-AT1R axis, often increasing ACE2 and Ang (1–7) significantly[21], in contrast with the inhibitory effects on the ACE2-ANG(1-7) axis by gluco-corticoids[51]. On the other hand, other substances are usually considered dangerous in some treatments for their negative effects on ACE2 levels.

The discovery of the role of ACE2 as the membrane receptor of SARS-CoVs and SARS-CoV2, raises controversial opinions concerning the effect of medical treatments in facilitating virus infection in human body. Because of the comorbidity of COVID19 with several diseases, like hypertension and diabetes[52], conventional pharmacological treatments are also under consideration. Hypotheses have been therefore put forward, especially from social media, to suggest potential adverse effects of angiotensin converting enzyme inhibitors (ACE-i) or Angiotensin Receptor Blockers (ARBs or Sartans), in COVID-2019, because of their action on the RAS pathway. This also provoked justified alarmed recommendations from different Scientific Societies[53] that underlined the lack of evidence against ACE-i or ARB medication (http://www.nephjc.com/news/covidace2). However, scientific evidence concerning the effects of pharmacological treatments acting on the RAS signalling system are available from animals[24,54,55,56] and also from human tissues treatments[46]. This is the reason why a careful analysis on previous results can now help to clarify possible relationships of these medical treatments in predisposing to COVID19.

**The model**

This effort aimed to depict the main pathways determining the infection of SARS-CoV19 in the human body and the molecular framework for its establishment. The molecular model that emerged represents the RAS signalling system components and the possible links with SARS-CoV19 infection, mediated by the transmembrane protein ACE2[1] (ACE2m) (Figure 1). Documented (solid lines) as well as hypothetical edges (dashed lines) in the network inferred from the public available scientific research are reported. The effects of ACE2 shedding[55] and/or internalization[56] which are mediated by AT1r, one of the two receptor of Angotensin II, are also shown. ACE2 shedding is determined by the cleavage of its ectodomain and its release in the plasma[55], producing free ACE2 (ACE2p). The alternative route represents ACE2 internalization[56], where the ectodomain is supposed to be imported into the cell, inhibiting its functioning outside of the membrane, and producing what is here indicated as the internalized ACE2 (ACE2i). Both processes are reported to be mediated by AT1r when not blocked with ARBs. In case of AT1r blockage, eventually by pharmacological treatments, these two routes are not working (Figure 2, panels b and c).

The model also includes the S-priming, which is required for the entry of SARS-CoV19 in human cells[4], and the action described for cytokins IL6 and TGF beta 1 in favouring or limiting the intracellular production of cathepsins, respectively[3]. The model represents the documented role of ACE2 levels in triggering the anti-inflammatory response, in case of high expression, or in inhibiting it[42,38]. The possible connections with lung severe damages are also shown[47].

**The data**

In figure 2, we summarize the effects of treatments on RAS molecules resulting from experimental studies on animal[24,54] and on human tissues[57]. Molecules levels in terms of concentration or activity are reported in Table 1.

|      | Ishyiama et al. 2004 | Ferrario et al. 2005 | | | Koka et al. 2007 | |
|------|---|---|---|---|---|---|
|      | losartan/omesartan | lisinopril | losartan | Combined | ANGII | losartan |
| ACE2 |  |  | UP | UP | DOWN | UP |
|      |  | Control-l |  |  |  |  |
| ace2 | UP | UP | UP | Control-l | DOWN | UP |
| ACE  |  |  |  |  | UP | DOWN |
| ace  | Control-l | Control-l | Control-l | Control-l | UP | DOWN |
| ANGII | UP | DOWN | UP | DOWN | UP | UP |

Table 1. Summary of results from exposure to ANGII and to ARBs and/or ACE inhibitors.[24,54,57]
Control-l are results comparable to the control. High (UP) and low (DOWN) levels based on the specific experimental measure are indicated.

Interestingly, the results here reported highlight that the treatments with sartans alone, or in combination with ACE inhibitors determine high activity of ACE2, while treatments with ACE inhibitors determine no changes in ACE2 activity.

Colours in Figure 2 highlight the differential levels of the molecules (black: corresponding to the control or not reported; red: high; blue: low). Solid red lines indicate the favoured routes that emerge in case of SARS-CoV19 presence. The possible pathways (red routes) are inferred based on molecule concentrations in the respective condition. In green boxes, clinical parameters to be collected and further investigated for disease progress monitoring, since their value could be affected when the viral infection is occurring. As an example, the monitoring of ANG(1-7) and ANG-II levels could reveal the activity of ACE2 in the RAS signalling pathways, together with ACE2p levels. CoV19/ACE2p (ACE2p in the plasma bound to virus) levels could further depict the extent of virus upload. As a result, high risk and low risk medical treatments as well as the disease development could be traced appropriately timely configuring individual predisposition and progress of clinical conditions.

**The clinical evidence**

This study was inspired and confirmed by one COVID19 case in Naples.

Sixty three years old man, under therapy for hypertension, treated with Reaptan (ACE-inhibitor). The dosage was set for an overweight man (105kg). The man was on diet during the last year till he reached 80Kg, while under the same medical treatment. Showing weakness, his dosage was reduced a couple of days before he discovered to be positive for COVID-19. When declared positive, the man was suggested to change his treatment using only Valsartan (Courtesy from Direct Personal Communication). During the disease, he manifested mild fever and nausea. The man fully recovered in 10 days and is now waiting for negative confirmatory tests.

**Hypothesis**
The long term, overdosage, ACE-inhibitor treatment limited the SARS-CoV2 virulence in the single case patient here reported for two main reasons, which are clearly depicted by the model. First, the virus attack was hampered by the reduced ACE2 availability known to be triggered by the medical treatment[54], that provoked a decrease of the number of viral S-domain receptors (ACE2m) on cell membranes, therefore limiting the viral entry in cells. Second, in case shedding[55] would have occurred, favoured by the availability of AT1r which is not blocked by other medical treatments, ACE2m is cleaved and released in the form of a soluble molecule (ACEp), thus increasing the amount of circulating SARS-COV2 receptor domain in the plasma. Because SARS-CoV2 replication can be blocked by soluble forms of ACE2[1], we hypothesize that ACE2p in plasma sequesters SARS-CoV2 and reduces the probability of viral internalization in human cells, conferring the additional protection from the viral infection. Proactive compounds based on the administering to individuals of free ACE2 as a medical treatment have already been proposed for limiting viral cellular invasion[58,44].
Hence, prolonged exclusive use of ACE-inhibitors and avoidance of AT1r blockade by ARBs (Sartans), that could be exerted by a combine treatment, could have provided a higher protection from the virus invasion for this patient. The later administering of medical treatments based on sartans, that was inappropriately suggested, fortunately did not affect a milder progress of the disease that favoured a successful recovery. This was highly probable due to an initial limited viral infection because of lack of ACE2 cell receptors conferred by the overdosage treatment based on ACE-inhibitors.
Monitoring of free ACE2p or of soluble ACE2 bound to viruses (ACE2v), together with ANG-II and ANG(1-7) levels and aldosterone in positive patients, can be the appropriate clinical trial that could further support the proposed hypothesis and be associated to current screening and monitoring for disease predisposition testing.
Unfortunately, the patient could not at all be monitored based on these parameters because of the emergency status in Italy.

**Discussion**
Many information to fight COVID19 can be inferred from the scientific efforts and public biological databases thanks to studies on previous epidemic coronaviruses.
The main effort here was to exploit this knowledge to define the SARS-CoV2 pathways when infecting the human body and the molecular bases for considering risk factors related to specific treatments.
Our molecular model elucidates the relationships between the RAS signalling and the SARS-CoV infection, including additional routes that can be inferred based on scientific evidence. The levels of the involved molecules were estimated considering six different experiments related to treatments for hypertension in both animals[24,54,55,56] and human tissues treatments[46]. These results, when associated with the model, highlight infection risk prone and protective medical interventions. In the case of ARBs (Sartans) and ACE inhibitor treatments, the model reveals that ARBs alone, or in combined treatments, that confer an increased ACE2 activity on the cell membranes, results in higher risk for viral infection. In contrast, ACE-inhibitors (when not combined with ARBs) appear to exert a two-way protective role, decreasing ACE2m on cell membranes and increasing ACE2p in the plasma, thus offering a possible alternative to medical treatments for hypertension. One evidence from a positive fast recovering patient is surely not enough to prove our hypothesis, but it was the evidence that inspired it, and this should encourage clinical trials and appropriate statistics that are still lacking and are urgently required. We are suggesting, as an example, the monitoring of RAS metabolites or of free or virus bound-ACE2 in the plasma, that could provide crucial information to answer questions on disease predisposition, on its progress and on mechanisms that are affected by the viral infection. Such approaches could be associated with the screening, the diagnostic, the monitoring and will surely highlight novel opportunities: as an example, in confirming the administering efficacy of ACE2 or ACE2-like molecules to prevent viral invasion in the organism. This data would also provide further evidence to convince the mainstream beliefs on the evidence here shown, for those who hardly accept model driven knowledge. The good news is that there are alternatives for hypertension treatments, and suitable medical guideline could be fast adopted based on recommendations of the pertinent medical societies.

Attention should also focus on monitoring COVID19 propensities for the other diseases or treatments that trigger ACE2 increase, therefore predisposing to the critical progress of the viral infection.[24,29,30,31,32,33,34,35,36] In parallel, careful studies should also consider COVID19 progress in cases in which a decreased ACE2 content occurs, ARDS included.[47,48] Beyond being supportive for the presented model, these studies could also reveal additional pathways of infection, expanding current knowledge and paving the way to novel investigations. Low ACE2 levels, in principle, depress the risk of a viral invasion since the depletion of viral receptor binding sites; on the other hand, one should keep in consideration that such predispositions were indeed revealed in occurrence of injuries triggered by the negative action on the ACE2-ANG(1-7) axis of the RAS signalling. Some questions therefore arise: who will play the major role in the viral attack? The decrease in viral receptor binding site, the presence of preceding injuries, or the medical care that was provided to overcome the injuries and the poor ACE2 content?

The World Health Organization communicated that the SARS typically attacks the lungs in three phases: viral replication, immune hyper-reactivity, and pulmonary destruction [https://www.nationalgeographic.com/science/2020/02/here-is-what-coronavirus-does-to-the-body/]. Scientific evidence, however, highlights that the distribution of the SARS-CoVs receptor (ACE2), does not strictly correlate with SARS-CoVs cell tropism in lungs. Expression patterns distribution in different organs and tissues of molecules triggering the viral entry[9,10,11] revealed that the virus could enter the cells through the tongue[12], thanks to the co-presence of ACE2 and of proteases favouring the S-priming, triggering the initial immune response. The release of specific cytokines such as IL6 could also cooperate to the S-priming[4,13]. Remarkably, IL6 and not TGF-beta1 appears to be overproduced in COVID-19 patients[14,15]. The evidence that the inflammatory response could also contribute to the viral entry highlights an intriguing cooperative mechanism and an alarming scenario on the viral mechanisms of infection and on its evolution.

The spreading of the infection to the lung cells could be through the vascular system, modulated by patient predisposition (i.e. ACE2 levels on cell membranes along the vascular endothelium)[11] and by the individual capability to mount the appropriate immune response. The initial lung disturb, presumably due to viral destruction of lung infected cilia cells, like in SARS, produce necrotic cells shed onto the luminal surface of the epithelium, contributing to a less reactive tissue response.[59]

The critical development of the disease depends on further additional conditions. Predisposing effects eventually due to previous injuries or inflammatory status, or to preceding pharmacological treatments, or to erroneous pharmacological treatments at the initial stage of infection[50,21], could facilitate the virus fast spreading, accompanied by an advantageous environment emerging from the side effects caused by its infection (for example ACE2 depletion with further injuries and increased inflammation). The starting infection in the mouth and the secondary attack in the lungs via the vascular transportation is also suggested by evidence of the early bilateral lung injuries, highlighted by computational tomography and radiography images.

The acute severe pneumonia is the result of both a depletion of ACE2 content on cell membranes, noticeable creating damages, and a delayed mounting of the adaptive immune response[60], contributing to alveoli engulfment due to an excessive innate response. This could furthermore justify the emergency from aged population, that could mount a slower response, putatively because of the reduced leukocytes levels. Subsequent organ failure could be facilitated by an initial higher ACE2m content in many key tissues[9,10,11], then depleted by the viral attack. Organ failure is also triggered by the infection mediated increase of dangerous components from the ACE-ANGII axis of the RAS signalling, with consequent cascades of events that cause typical effects like cardiovascular and kidney injuries. Kidney has a high content of ACE2 in physiological conditions and the renin/aldosterone branch can be the part of the RAS system playing a role in the organ failure, also linking aldosterone levels to disease propensity. This also explains why men are more disease prone than women[61], as it comes evident from the number of death on number of cases per sex (male: 4.7%; female: 2.8%), as reported in the disease statistics on March 23, 2020
[https://www.worldometers.info/coronavirus/coronavirus-age-sex-demographics/].
All concurrent events to be added to the severe cytokine storm that is damaging the lung tissues causing hypoxia.
Hypoxia is the main aspect today faced during the emergency in the hospitals. The medical intervention against the cytokines storm to reduce the acute respiratory distress syndrome need to be carefully targeted to avoid immune depression and consequent infection by secondary pathogens, although the immune depression is not a predisposing factor for this disease. Attention should also be focused on avoiding medical treatments that could increase ACE2m content with the aim to decrease inflammation[50,21].
All the relevant molecular connections are elucidated in the proposed model, accompanied, by the depicting of the possible routes for the infection, based on scientific evidence and the analysis of the expression pattern per tissue of the main proteins involved in the viral entry from public biological databases.
What this research is also teaching is that, although a severe pandemy was expected, useful information was available, possible preparative actions were suggested at least since 2005[62], we were not ready!
One example comes from the evidence that the infection is connected with the RAS signalling and that there is still lack of appropriate trials to investigate on hypertension parameters for risk assessment and disease progress monitoring. The lack of public digitalized data from diagnostics in the hospitals, and from the medical surveillance at home, on COVID19 positive patients that do not manifest a severe disease is also evident. Shared tools for diagnostics and coordinated intervention should be fixed soon. The emergency must be faced with proper instruments.
What can we do now?
While working to fight the disease with vaccine development and adopting new or re-purposed antiviral medicines to limit the virus attack, we must also use all the knowledge at our disposal to treat those patients most at risk of dying, collecting other data for further understanding the disease. This includes primarily to urgently implement evidence-based and science driven approaches that could efficiently accompany those who are on the front line in the hospitals and are facing the emergency. The implementation of diagnostics methods, medical treatments and tools based on the monitoring of the individual clinical conditions, and the set-up of appropriate intervention for targeted groups of patients, should profit of a joint action with a proactive scientific community and the support of stakeholders.
We really hope that this alerting will be foster these actions.

**Additional information**
In Table 2 we report the take home messages that were gained from the presented analysis COVID19. The model is presented in a dedicated website in the framework of "LETS'BE: let's be ready!", a transdisciplinary initiative to make scientific knowledge understandable and accessible to the interested community, to be conscious for the current fight and be prepared for the next.
For further information and possible interest in giving a contribution, please subscribe to http://bigdatainhealth.org/LETSBE.

v2: this version does not include Table 1 of the previous version which was copied from
http://www.nephjc.com/news/covidace2.

**Acknowledgments**
I wish to acknowledge:
the positive COVID19 patient in Naples that kindly provided information useful to inspire the work. He is still waiting for the test to confirm complete recovery, because of the emergency focused on more urgent issues. It would have been of interest to have a time course monitoring of clinical parameters from him during the disease and soon after, for more understanding;
Dr. Antonino Chiusano (radiologist) for useful explanations on data from public radiography images.
Dr. Antonio Scala (physicist) for his warm support and for useful suggestions;
Dr. Clementina Sansone (biologist) for the evening she spent for a scientific discussion on viruses;
Prof. Stefano Mazzoleni for inspiring in me my model based approach for understanding biology.

I wish to apologize with my family, for my absence, although being at home because of SARS-CoV2. Hoping that my full day effort in these days will be never for them.

**Table 2. Summary of the main evidence derived from the work.**

-The presence of ACE2 and associated proteins favouring the viral entry in oral tissues suggests that the initial infection could arise preferentially in the mouth and, if favoured by clinical conditions and not arrested by an efficient immune response, would spread to the lung through the vascular system, provoking a cytokine storm that can be fatal.

-An over infection in lung tissues, mediated by previous or concurrent medical treatments or predisposition might enhance fast engulfment of the alveoli due to an over response of the innate immune system. This is accompanied by the cooperative effect of the misfunctioning of the ACE2-ANG(1-7) axis of the RAS signalling system, that further contributes to the inflammatory response, while not anymore protecting lungs from injuries.

-Increase of ACE2 content could favour viral invasion (evidence of ACE2 increase are found in scientific literature for mineralocorticoid receptor blockers (MRB), ibuprofen, ARBs-Sartans).

-Substances could sequester or inactivate molecules that reduce ACE2 content; as an example, AT1r sequestering by ARBs (Sartans) increases ACE2 content on cell membranes.

-Substances could favour shedding or internalization of ACE2m could decrease the viral infection rate, although side effects due to the decreased functioning of the ACE2-ANG(1-7) RAS signalling axis.

-Exclusive treatments by ACE-inhibitors (not accompanied by ARBs (Sartan) treatments) could exert a protective effect.

-Targeted actions based on deep diagnostics to reduce cytokine overload (not using general anti-inflammatory treatments) should be provided since some cytokines appears to favour while others inhibit the viral infection (examples are reported in model here provided (Figure 1)).

-Information on cytokines average content from COVID-19 patients (critical or dead) are available from Wuhan patients and from SARS-CoV previous pandemic. This diagnostic parameters could be useful to be traced per patient, and could provide relevant information for treatments and for further investigations on the disease progression.

-Inappropriate treatments could favour attacks from secondary pathogens (eg. decreasing cytokins)

-Inappropriate treatments could favour the viral invasion (eg. increasing ACE2 content)

- ACE2 levels increase in a number of pathologies, as here reviewed, and that predispose to the infection.

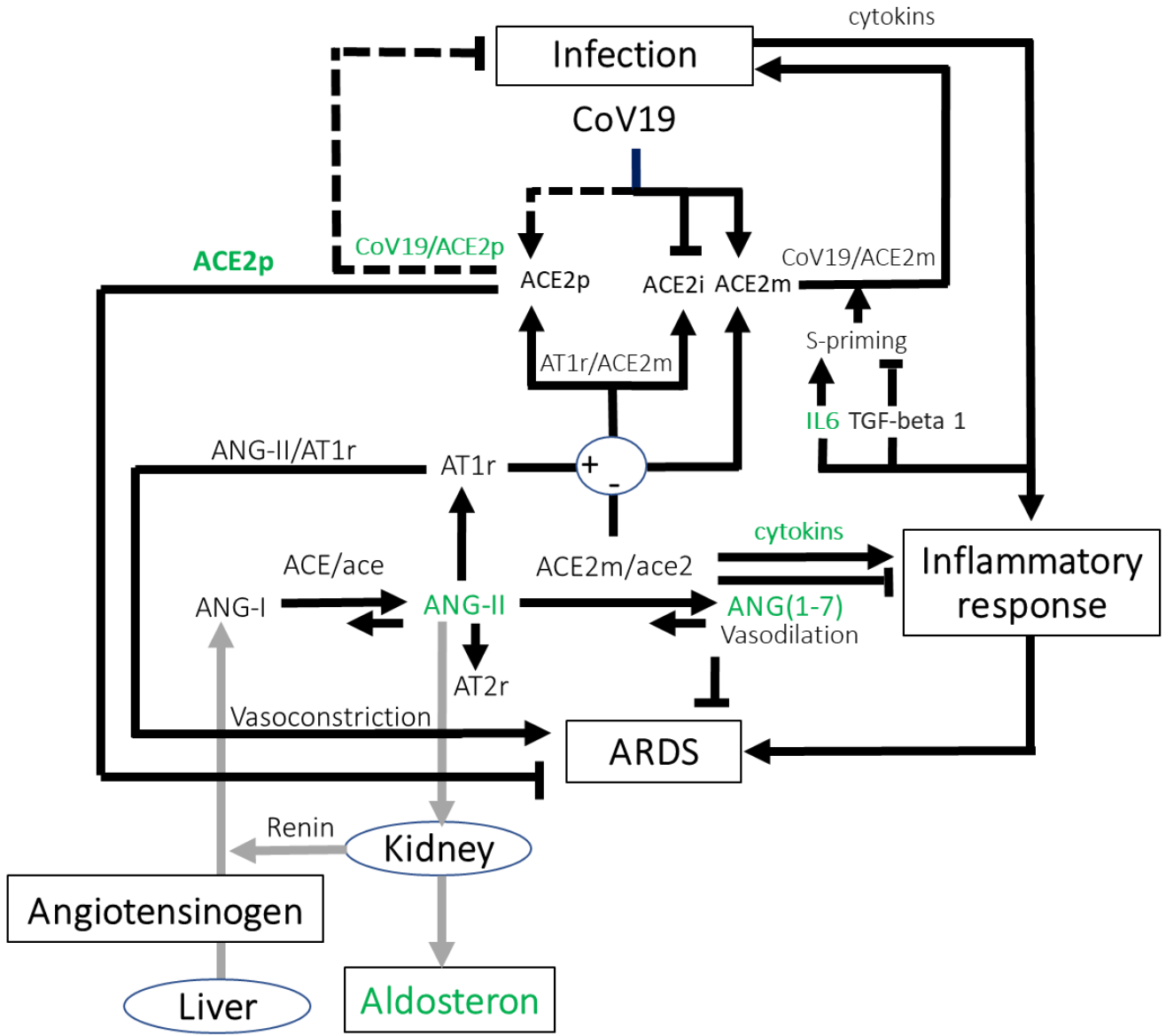

Figure 2. Main components of RAS and of known pathways from viral (CoV19) infection. Solid lines represent known connections. Dashed lines represent inferred connections based on scientific evidence. In green parameters that would need monitoring in COVID19 infected patients for risk assessment.

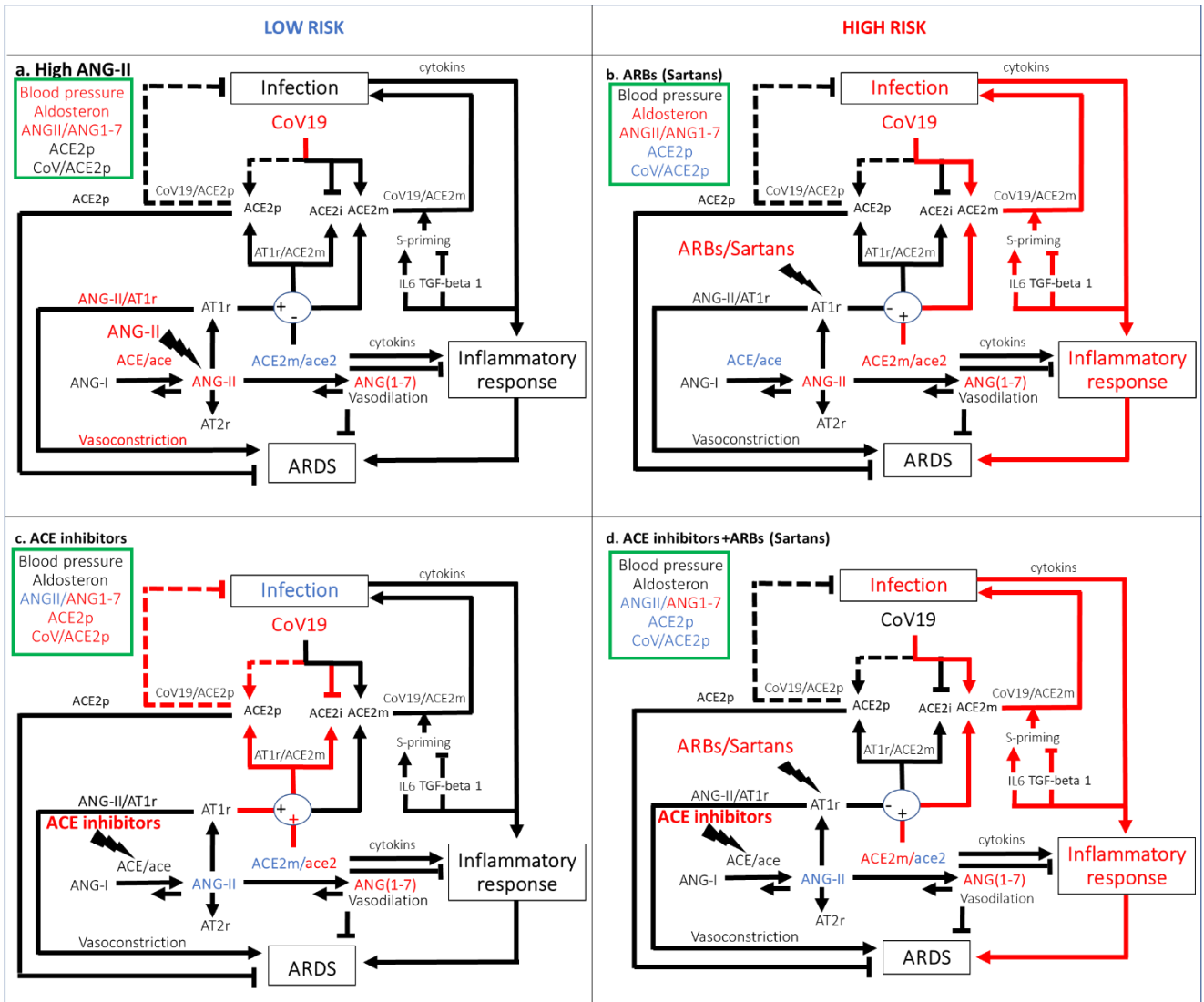

Figure 2. Possible routes for viral infection according to medical treatments affecting the RAS signalling (panels b, c, d). Panel a indicates high ANG-II content conditions (eg. in hypertension). High risk and low risk treatments to viral infection are shown.

a. Results from in vitro experiments on human tubular cell lines (HK-2) which were performed to analyse the response to increasing dosages of Ang II[57] show up-regulation of ACE mRNA (ace) as well as of the encoded protein (ACE), while reducing ace2 and ACE2m content, respectively.

b. Exposure to losartan, an AT1r blocker (ARB), resulted in decreased ACE levels for both mRNAs and proteins and in the increase of ACE2 (mRNA and protein).

Results from treatments with sartans on rats[54] revealed increased ACE2 and ace2, indicating high ACE2 activity and high mRNA levels, in agreement with what reported for human cell lines[57] for similar treatments (b).

c. Increased mRNA levels and control-like ACE2 activity were revealed under exposure to lisinopril (an ACE inhibitor).

d. Control like ace2 mRNA expression and an increased ACE2 activity resulted in the combination of the two treatments. Noticeable, in rats medicated with lisinopril, increased cardiac ace2 gene transcription did not elicit a comparable increase in ACE2 activity.

In green boxes, some of the parameters that would need monitoring for appropriate health and risk assessments in individuals.